\documentclass[sigconf]{acmart}

\usepackage{comment}
\usepackage{framed}
\usepackage[framemethod=tikz]{mdframed}
\definecolor{shadecolor}{rgb}{0.75, 0.75, 0.75}
\definecolor{light-gray}{gray}{0.95}
\usepackage{float}
\usepackage{booktabs}
\usepackage{longtable}
\usepackage{soul}
\usepackage{fixltx2e}
\usepackage{flushend}

\newcommand{\rqone}{Why do Brazilian SE researchers use grey literature?}
\newcommand{\rqtwo}{What types of grey literature are used by Brazilian SE researchers?}
\newcommand{\rqthree}{What are the criteria Brazilian SE researchers employ to assess grey literature credibility?}
\newcommand{\rqfour}{What benefits and challenges Brazilian SE researchers perceive when using grey literature?}

\AtBeginDocument{%
  \providecommand\BibTeX{{%
    \normalfont B\kern-0.5em{\scshape i\kern-0.25em b}\kern-0.8em\TeX}}}

\copyrightyear{2020}
\acmYear{2020}
\setcopyright{othergov}\acmConference[SBES '20]{34th Brazilian Symposium on Software Engineering}{October 21--23, 2020}{Natal, Brazil}
\acmBooktitle{34th Brazilian Symposium on Software Engineering (SBES '20), October 21--23, 2020, Natal, Brazil}
\acmPrice{15.00}
\acmDOI{10.1145/3422392.3422442}
\acmISBN{978-1-4503-8753-8/20/09}

\begin{document}






\title{On the Use of Grey Literature: A Survey with the Brazilian Software Engineering Research Community}


\author{Fernando Kamei}
\orcid{0000-0002-5572-2049}
\affiliation{%
   \institution{UFPE, IFAL}
  \city{Macei\'o}
  \state{Alagoas}
  \country{Brazil}}
\email{fernando.kenji@ifal.edu.br}

\author{Igor Wiese}
\affiliation{%
  \institution{UTFPR}
  \city{Campo Mour\~ao}
  \state{Paran\'a}
  \country{Brazil}}
\email{igor@utfpr.edu.br}

\author{Gustavo Pinto}
\affiliation{%
  \institution{UFPA}
  \city{Bel\'em}
  \state{Par\'a}
  \country{Brazil}}
\email{gpinto@ufpa.br}

\author{M\'arcio Ribeiro}
\affiliation{%
  \institution{UFAL}
  \city{Macei\'o}
  \state{Alagoas}
  \country{Brazil}}
\email{marcio@ic.ufal.br}

\author{S\'ergio Soares}
\affiliation{%
  \institution{UFPE}
  \city{Recife}
  \state{Pernambuco}
  \country{Brazil}}
\email{scbs@cin.ufpe.br}

\renewcommand{\shortauthors}{Kamei et al.}

\begin{abstract}
    \textbf{Background:} The use of Grey Literature (GL) has been investigated in diverse research areas. In Software Engineering (SE), this topic has an increasing interest over the last years. 
    \textbf{Problem:} Even with the increase of GL published in diverse sources, the understanding of their use on the SE research community is still controversial. 
    \textbf{Objective:} To understand how Brazilian SE researchers use GL, we aimed to become aware of the criteria to assess the credibility of their use, as well as the benefits and challenges. 
    \textbf{Method:} We surveyed 76 active SE researchers participants of a flagship SE conference in Brazil, using a questionnaire with 11 questions to share their views on the use of GL in the context of SE research. We followed a qualitative approach to analyze open questions.  
    \textbf{Results:} We found that most surveyed researchers use GL mainly to understand new topics. Our work identified new findings, including: 1) GL sources used by SE researchers (e.g., blogs, community website); 2) motivations to use (e.g., to understand problems and to complement research findings) or reasons to avoid GL (e.g., lack of reliability, lack of scientific value); 3) the benefit that is easy to access and read GL and the challenge of GL to have its scientific value recognized; and 4) criteria to assess GL credibility, showing the importance of the content owner to be renowned (e.g., renowned author and institutions). 
    \textbf{Conclusions:} Our findings contribute to form a body of knowledge on the use of GL by SE researchers, by discussing novel (some contradictory) results and providing a set of lessons learned to both SE researchers and practitioners. 
\end{abstract}


\maketitle

\section{Introduction}
Grey Literature (GL) is a data source that was not subjected to quality control mechanisms (peer review) before publication~\cite{Petticrew:2006:Book:SR}. Several areas of knowledge investigated the use of GL, for instance, Medicine~\cite{Paez:2017:Medicine} and Management~\cite{Adams:2017:Shades:IJMR}. According to Paez~\cite{Paez:2017:Medicine}, GL may provide data not found within commercially published literature, providing an important forum for disseminating studies with null or negative results that might not otherwise be disseminated, which in turn reduce publication bias to the propensity for only studies reporting positive findings to be published, increase reviews’ comprehensiveness and timeliness, and foster a balanced picture of available evidence.

In the context of Software Engineering (SE) research, there is an increasing interest in the investigation about GL over the last years. This was particularly motivated due to the also growing mediums that SE practitioners use to exchange problems and ideas, including news aggregator websites such as Reddit and Hacker News~\cite{Aniche:ICSE;2018}  
and question and answer (Q\&A) websites such as Stack Overflow~\cite{Zahedi:2020:EASE}. 
Although studies recognized the importance and usefulness of the GL in general~\cite{Garousi:2016:EASE}, and blog content, in particular~\cite{Williams:2017:EASE, Williams:2018:ASWEC}, there is a lack of understanding on how to properly use GL, (for instance, how to find acceptable evidence),
which brings challenges for researchers that are interested in using this kind of medium in their research.


The goal of this research is to investigate the perceptions of Brazilian SE researchers on the use of GL. This research is important to the SE research community to improve the understanding of how researchers could explore and take advantage of GL, approximating their research findings to practice. Still, the content provided by software practitioners, if created with rigor and quality, could be useful to advance the state of the art.

To achieve this goal, we explored four research questions:
\begin{itemize}
    \item \textbf{RQ1:} \textit{\rqone}
    \item \textbf{RQ2:} \textit{\rqtwo} 
     \item \textbf{RQ3:} \textit{\rqthree} 
     \item \textbf{RQ4:} \textit{\rqfour} 
\end{itemize}

Answering \textbf{RQ1} and \textbf{RQ2} is essential to understand to what extent the Brazilian research community use GL, and what motivates this community to use it. To the best of our knowledge, this is the first study that investigated Grey Literature with the Brazilian SE research community. Answering \textbf{RQ3} is essential to understand the reliability criteria that will be important to researchers to better select evidence retrieved from a GL source, and for practitioners to better understand on how to increase the credibility of their content shared. Finally, answering \textbf{RQ4} is important to understand the potential benefits and challenges using GL more broadly, by researchers and practitioners. 


To answer these research questions, we surveyed 76 Brazilian SE researchers. The evidence obtained from a qualitative analysis of the answers yields important lessons that can inspire SE researchers and practitioners who investigate and provide content in a diversity of GL source. 

In summary, in our work: 1) we elucidate the first perceptions about GL in Brazilian SE community; 2) we found the main GL sources used by Brazilian SE researchers; 3) we noted several motivations to use or reasons to avoid GL, highlighting the importance to better investigate how researchers and practitioners should deal with GL; 4) we provided different perspectives to assess GL source credibility from previous studies, showing the importance of being a renowned author; 5) we provided important advice with lessons learned on how to deal with GL, to both researchers and practitioners; and 6) we confirmed previous findings and complement the state of art with new findings. 



\section{Background}\label{sec:background}

The term ``Grey Literature'' (GL) has many definitions, but the most widely accepted is the Luxembourg definition~\cite{Garousi:2019:IST}, 
that states: \textit{``[GL] is produced on all levels of government, academics, business and industry in print and electronic formats, but which is not controlled by commercial publishers, i.e., where publishing is not the primary activity of the producing body''.} In summary, the term ``grey'' 
literature is often used to refer to the literature that is not 
subject of quality control mechanisms (e.g., peer review) before a publication~\cite{Petticrew:2006:Book:SR}.

Adams et al.~\cite{Adams:2016:Health} introduce the idea of ``grey information'' to distinguish different forms of grey, including grey literature, grey information, and grey data. The term ``grey data'' is used to describe user-generated web-content, e.g., tweets, blogs. On the other hand, ``grey information'' is informally published or not published at all, e.g., meeting notes, emails~\cite{Rainer:2018:TR}. However, the SE literature hardly distinguishes these terms. Similarly, in our work, we considered all forms of grey data and grey information as GL. 

The use of GL in other disciplines is not recent. The maturity is perceived, for instance, by the support of GL research through several GL databases, repositories, and search engines (e.g., GreyNet\footnote{www.greynet.org}, OpenGrey\footnote{www.opengrey.eu}).
Moreover, there are several guidelines to support researchers to conduct a Grey Literature Review (GLR), such as management~\cite{Adams:2017:Shades:IJMR} and medicine~\cite{Paez:2017:Medicine}.

\begin{figure}[h!]
\includegraphics[scale = 0.27, clip = true, trim= 0px 0px 0px 0px]{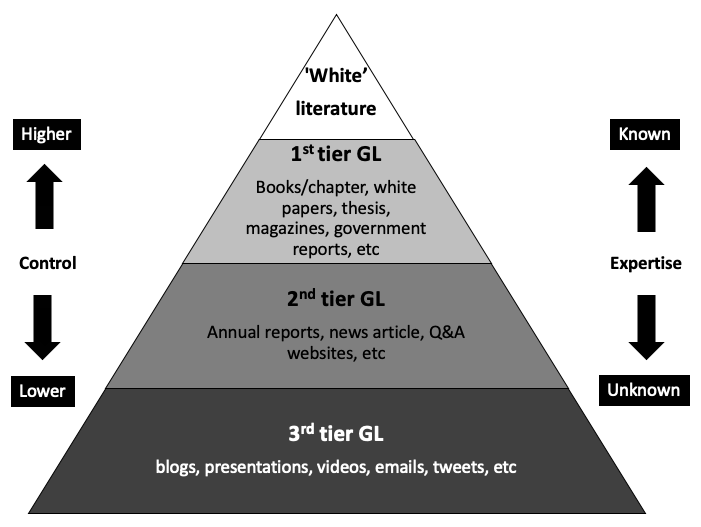}
\caption{The ``shades'' of grey literature, adapted of Garousi~\cite{Garousi:2019:IST}.}
\label{fig:shades-gl}
\end{figure}

GL has a wide variety of types that vary in the type of information that produces. Adams et al.~\cite{Adams:2017:Shades:IJMR} classified them according to ``shades of grey''. In the SE, Garousi et al.~\cite{Garousi:2019:IST} adapted these shades according to three-tiers (see Figure~\ref{fig:shades-gl}). These tiers running according to two dimensions: expertise and outlet control. Both dimensions run between extremes ``unknown'' and ``known''. The darker the color, the less moderated or edited the source in conformance with explicit and transparent knowledge creation criteria. 

In the context of SE, researchers have been using GL for several purposes. Some primary studies were conducted relying (mostly or entirely) on GL available on practitioners mediums, for instance, Stack Overflow~\cite{Zahedi:2020:EASE} and HackerNews~\cite{Aniche:ICSE;2018}. We also found several secondary studies that were conducted using GL, for instance, the SLR of Selleri et al.~\cite{Selleri:2015} that investigated the use of Agile methods with CMMI and were included some technical reports as primary studies. Moreover, there several Mapping Studies, for instance, the study of Sharma and Spinellis~\cite{Sharma:2018} that included some books as a reference to investigate knowledge related to software smells and identify challenges as well as opportunities. 

When GL is used as part of an SLR, it is called a Multivocal Literature Review (MLR). The term ``multivocal'' refers to diverse types of the source to be included as literature (white literature and grey literature). Note that MLR does not force researchers to use only GL. Instead, researchers can complement the findings of a traditional SLR with data from the GL. Kitchenham et al.~\cite{Kitchenham:2009:ESEM} conducted the first MLR in SE. This research aimed to compare the use of manual and automated searches and to assess the importance and the breadth of GL. Their findings showed the importance of GL, especially to investigate research questions that need practical and technical answers. However, it was observed that the quality of GL studies was lower than papers published in conferences and journals due to the criteria of quality assessment used that increase the grade for peer-reviewed studies. 

Garousi et al.~\cite{Garousi:2016:EASE} observed the importance of including GL to strengthen the evidence derived from practitioners when compared to the differences between the outcomes of an SLR and an MLR. After, Garousi et al.~\cite{Garousi:2019:IST} proposed a guideline to conduct MLR in SE. Another type of secondary study is the GLR that uses only GL as a source of primary studies. In SE, a GLR study was recently conducted by Raulamo-Jurvanen et al.~\cite{Raulamo-Jurvanen:2017:EASE}, which used only GL as based content. This GLR intends to understand how practitioners tackle the problem of choosing the right test automation tool. Their findings showed that practitioners tend to have a general interest in and be influenced by related GL.

GL is also used in SE tertiary studies, as we found in three studies that focused on GL. The first study~\cite{Yasin:Thesis:2020} investigated the evidence of GL use in the synthesis of secondary studies, showing that GL was present in around 9\% of SLRs synthesis discussion. The second study~\cite{Neto:2019:ESEM} investigated the motivations of SE researchers of 12 secondary studies had to conduct MLR, showing that MLR use was in the early stages. Their findings also showed motivations to conduct MLR, such as lack of academic research on the topic, the evidence available in GL, and when the research topic is emerging. The third study was recently published by Zhang et al.~\cite{Zhang:2020:ICSE} that found a group of 102 SE secondary studies that included GL as primary studies. Their findings showed the technical report as the most common GL type used in the studies, followed by white papers, blogs, book/book chapters, and thesis, respectively. Still, were investigated the motivations and challenges to use GL by surveying SE researchers.

\section{Research Questions}\label{sec:rqs}

In this work, we have four research questions. After stating the research questions, we describe a rationale for their purpose.

\vspace{0.2cm}
\begin{mdframed}
    \textbf{RQ1:} \textit{\rqone}
\end{mdframed}
\vspace{0.2cm}

\noindent
\textit{Rationale:} The widespread presence of GL mediums is posing a challenge for researchers. On the one hand, SE researchers could take advantage of GL to expand their notion of how developers use tools, solve their problems, or find knowledge. On the other hand, the non-peer-review nature of GL could make researchers skeptical about their credibility. Although some researchers may be inclined to use GL in their research, others may not.
In this broad question, we intend to understand if Brazilian SE researchers are using GL and, if so, what motivates them to use, or if not, the reasons that lead to not use GL. 

\vspace{0.2cm}
\begin{mdframed}
    \textbf{RQ2:} \textit{\rqtwo}
\end{mdframed}
\vspace{0.2cm}

\noindent
\textit{Rationale:} Nowadays, GL is available in many forms, from traditional mediums such as blogs, and question \& answer websites, to more dynamic mediums such as Slack and Telegram, to videos on YouTube, to interactive gaming discussions on Twitch. Each one of these forums offers researchers a rich spectrum of unstructured data, which could bring specific benefits and limitations. In this research question, we sought to investigate what sources are often used by Brazilian SE researchers. 
A better understanding of the GL source would be important to guide future research in this area. 

\vspace{0.2cm}
\begin{mdframed}
    \textbf{RQ3:} \textit{\rqthree}
\end{mdframed}
\vspace{0.2cm}

\noindent
\textit{Rationale:} GL is, by nature, not peer-reviewed; that is, when writing a blog post, practitioners are free to share their thoughts without worrying too much about methodological concerns. This freedom, however, may come with a cost: GL may be inaccurate, lacking context or details, or may even be incorrect. For instance, Fischer et al.~\cite{Fischer:2017:SP} analyzed 1.3 million Android applications and 15.4\% of them contained security-related code snippets from Stack Overflow. Out of this, 97.9\% contain at least one insecure code snippet. Therefore, when using GL in research works, researchers should employ additional levels of assessment to make sure the selected GL is indeed appropriate for the study. Answering this question will help us to understand the reliability criteria that Brazilian SE researchers consider. 

\vspace{0.2cm}
\begin{mdframed}
    \textbf{RQ4:} \textit{\rqfour}
\end{mdframed}
\vspace{0.2cm}

\noindent
\textit{Rationale:} Over the years SE researchers increased their interest in GL because some of them provide information from the SE practice, which is important to improve the research and fill the gaps. For instance, Zahedi et al.~\cite{Zahedi:2020:EASE} explored the Stack Overflow and found some trends and challenges in continuous SE that researchers could better explore. On the other hand, this understanding may be biased and with a lack of contextual or information. In this question, we are interested in understanding, in greater detail, the benefits and challenges that researchers may face when resorting to GL. 


To answer these questions, we employed mostly qualitative methods. In what follows, we present our survey instrument along with the procedures to collect and analyze the data.

\section{Research Method: A Survey}\label{sec:survey}
In this work, we focused on SE researchers potentially interested in using GL in their works. We followed the guideline of Lin{\aa}ker et al.~\cite{Linaker:2015:SurveyGuideline}, aiming to use a survey methodology to collect information from a group of people by sampling individuals from a large population. In this section, we describe the subjects (Section~\ref{sec:method-survey-subjects}), the questions of our questionnaire (Section~\ref{sec:method-survey-questions}), and the procedure we employed to analyze data (Section~\ref{sec:method-survey-analysis}).

\subsection{Survey Subjects}\label{sec:method-survey-subjects}

Our population comprehends SE researchers potentially interested in using GL in their research. We chose our sample using non-probabilistic sampling by convenience. Our sample comprehends participants of 
The Brazilian Conference on Software: Practice and Theory (CBSoft), the largest Brazilian software conference with the participation of many SE researchers and includes well-established and specialized satellite SE conferences in its domain.

We used two approaches to invite the researchers to answer our questionnaire. First, we placed posters on the walls and tables of the event with a brief description of the work and the link to the online survey. Second, we get the email addresses of the 252 participants. We asked the general chair of the conference whether s/he could share this information with us, which s/he gently provided. In our invitation email, we highlighted that the participant was attending the conference, and in the survey, we mentioned that the participant was free to withdraw at any moment, and all information stored was confidential. Before sending the actual survey, a draft survey was reviewed by an experienced researcher (PhD SE researcher with more then 15 years of experience in research). We also conducted a pilot survey. In this case, we randomly selected two participants and explicitly asked their feedback.
We received feedback suggesting to change the order of some questions and to re-write some questions to make them more understandable to the target population.
After employing these recommendations, we send the actual survey to the 250 remaining participants of the event.
In the invitation email, we briefly introduced ourselves, the purposes of the research, and the link to the online survey.
The survey was open for responses from September 26th to October 11th, 2019.
During this period, we received a total of 76 valid answers (30.4\% response rate). We did not consider the pilot survey answers.

Among the survey respondents, 48.7\% have a Ph.D., 31.6\% a Master's, 2.6\% are graduate specialization, 14.5\% a Bachelor's degree, and 2.6\% are undergraduate students. Among them, 72.4\% are male and 27.6\% are female.

\subsection{Survey Questions}\label{sec:method-survey-questions}

Our survey had 11 questions (only one was required, eight of which were open). For replication purposes, all the data used in this study is available online at: \texttt{https://bit.ly/31OUaYo}. 
We used different survey questions flow for those who have used GL (just did not answer question 10) from those who have not (answered only questions 1 to 4 and questions 10 and 11). The questions covered in the survey were:

\begin{enumerate}
    \item What is your e-mail? \{Open\}
    \item What is your gender? Choices: \{male, female, other\}
    \item Please list the highest academic degree you have received. Choices: \{High school, Technical education, University graduate, Expert, Master's degree, Doctorate\}
    \item Have you used grey literature? If you never used, go to question \textit{ten}. Choices: \{Yes, No\} \{Required*\}  \{RQ1\}
    \item What sources of grey literature did you use? \{Open\} \{RQ2\}
    \item In which conditions do you use grey literature? \{Open\} \{RQ1\}
    \item In which conditions do you do not use grey literature? \{Open\} \{RQ1\}
    \item Could you list any benefits in using grey literature? \{Open\} \{RQ4\}
    \item Could you list any challenges in using grey literature? \{Open\} \{RQ4\}
    \item If you answered no in the question \textit{four}, please state why did you never use or avoid use grey literature? \{Open\} \{RQ3\}
    \item What would be a reliable source of grey literature for you? \{Open\} \{RQ3\}
\end{enumerate}

\subsection{Survey Analysis}\label{sec:method-survey-analysis}

Two independent SE researchers, a Ph.D. student and a Ph.D. professor, both with previous experience in conducting qualitative research, followed qualitative procedures to extract and analyze the questionnaire data. 

We performed an agreement analysis with the codes and categories generated by each researcher using the Kappa statistic~\cite{viera2005}. The Kappa value was 0.749, which means a Substantial Agreement level, according to the Kappa reference table~\cite{viera2005}. We then detail the procedure used to analyze the answers (adopted and adapted from~\cite{Pinto:ICSE-SEET:2019}).

\begin{enumerate}
   \item \emph{Familiarizing with data:} The answers of the survey respondents were read by the two independent researchers.
   \item \emph{Initial coding:} In this step, we individually added codes. We used a post-formed code, so we labeled portions of text without any previous pre-formed code, that is, labels that could express the meaning of the excerpts of the answer that had appropriate actions or perceptions. The initial codes are considered temporaries since they still need refinement. The codes were identified and refined throughout all the analysis. An example of coding is present in Figure~\ref{fig:labeled-text}.

    \begin{figure}[h!]
    \includegraphics[scale = 0.45, clip = true, trim= 0px 0px 0px 0px]{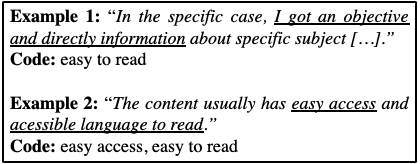}
    \caption{Coding process used in a questionnaire answer.}
    \label{fig:labeled-text}
    \end{figure}

   \item \emph{From codes to categories:} Here, we already had an initial list of codes. We then begin to look for similar codes in the data. We grouped the codes with similar characteristics in broader categories. Eventually, we also had to refine the categories found, comparing, and re-analyzing in parallel, using an approach similar to axial coding~\cite{Spencer2009}. An example of this process is presented in Figure~\ref{fig:codes-category}.

    \begin{figure}[h!]
    \includegraphics[scale = 0.45, clip = true, trim= 0px 0px 0px 0px]{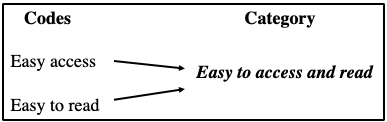}
    \caption{Example of how the category emerged from the initial codes.}
    \label{fig:codes-category}
    \end{figure}

    \item \emph{Categories refinement:} Here, we have a potential set of categories. We then, in consensus meeting, evaluated and solved the disagreements of interpretation for evidence that supported or refuted the categories found. We also rename or regroup some categories to describe the excerpts better there. Still, we invited a third researcher (a Ph.D. professor) to review and comment on those categories, and in case of any doubt, was started a discussion between the first two researchers.
 \end{enumerate}

\section{Results}\label{sec:results}

In this section, we present our main results organized in terms of the research questions. To enable traceability, we include direct quotes from respondents along with the answer identified in open-ended questions and we present the discovered codes slanted. We also presented the list of categories found in tables with the total number of occurrences of a given category in the column ``\#''. An important observation is that some researchers may have reported more than one answer per question, which may happen to be grouped into different categories.
Still, most of our questions are not required. Then, when summarizing the categories in tables, the overall results might not reach 100\% of respondents.

\subsection*{RQ1: \rqone}

\textbf{\textit{Overall use.}} Most of the respondents of our work (53/76 occurrences, 69.7\%) are using GL to some purpose, that means they had previous experience using GL. This value was used to analyze all the categories about motivations to use GL in the following. 
Moreover, to better understand why and how SE researchers are using GL, we asked questions that included the motivations to use GL or reasons to avoid it. We observed several driving motivations \textit{to use GL}, as present in Table~\ref{tab:motivations-use-GL}. We describe some of them in the following.

We highlight that one answer of researchers could be related to more than one category found. This is worth to the next following categories of other RQs.

\begin{table}[H]
\caption{Motivations to use GL.}
\label{tab:motivations-use-GL}
\begin{minipage}{.4\textwidth}
  \centering
  \begin{tabular}{lcc}
    \toprule
    Motivation & \# & \%\\
    \midrule
    To understand the problems & 28 & 52.8\%\\
    To complement research findings & 12 & 22.6\%\\
    To answer practical and technical questions & 10 & 18.9\%\\
    To prepare classes & 4 & 7.5\%\\
    To conduct government studies & 1 & 1.9\%\\ 
    \bottomrule
  \end{tabular}
\end{minipage}
\end{table}

\subsection*{Motivations to use}

\textbf{\textit{To understand the problems}} (28/53 occurrences, 52.8\%). This category was mentioned by more than half of respondents, which means when the researcher uses GL to understand or investigate a new topic that has no previous knowledge, or when s/he looks for something aiming to solve problems, or when they want to acquire specific information to deepen the knowledge. Regarding this category, some respondents have pointed out: \textit{``I used initially \ul{to learn a topic} that I don't have knowledge''}, \textit{``In most cases, \ul{to understand how the problem happens} in the society''}, and \textit{``When I want to \ul{search for deep references} and in a large amount about a specific theme.''}


\textbf{\textit{To complement research findings}} (12/53 occurrences, 22.6\%). A researcher mentioned that used GL to complement a Mapping Study, as we quoted out: \textit{``GL was used \ul{to complement a data of a Systematic Mapping}.''} Another respondent raised using GL for a specific context, in which the peer-reviewed content is still scarce, as pointed out: \textit{``I use it \ul{when I don't find many studies in a specific context}, for instance, in the use of SE in the context of digital games \ul{there are process models that are not described in articles that are considered by game developers}.''}

\textbf{\textit{To answer practical and technical questions}} (10/53 occurrences, 18.9\%). This category was quite mentioned, mainly about understanding the state of the practice. In this sense, a respondent pointed out: \textit{``(...) I use it when I have the perception that \ul{the theme has an origin on the industry} and is on \ul{discussion or an increase of adoption in the industry}.''}


\textbf{\textit{To prepare classes}} (4/53 occurrences, 7.5\%). Few SE researchers mentioned the use of GL to support the material to prepare classes, as a respondent pointed out: \textit{``(Use GL) When I'm searching for something for a class.''} In the investigated research community it is common for SE researchers are also professors at universities. For this reason, some researchers have used the GL to support them.



\subsection*{Reasons to avoid/never use} 

Even though the perception of several motivations to use GL, 50.9\% of SE researchers (27/53) \textbf{\textit{avoid using}} GL as a reference or to reinforce some claims in \textbf{Scientific papers}, or any other type of scientific documents, such as thesis and SLR, because they argued that evidence of GL is usually scarce of scientific value that makes it is not often well-regarded research community. In this regard, a SE researcher mentioned: \textit{``\ul{I try to avoid the use of GL in research papers and systematic reviews}. Generally, the community belittles such references.''} 
Furthermore, we found some respondents that \textbf{\textit{never used GL}} (23/76 occurrences, 30.3\%), that means they did not have previous experience using GL to any research situations. This value was used to analyze all the categories about reasons to never used GL in the following. Of those 23, 15 answered our question that intended to understand the reason to never use GL. The summary of the findings for this question is presented in Table~\ref{tab:never-used-GL}. We describe some of them in the following.

\begin{table}[tp]
\caption{Reasons to never use GL.}
\label{tab:never-used-GL}
\begin{minipage}{.4\textwidth}
  \centering
  \begin{tabular}{lcc}
    \toprule
    Motivation & \# & \%\\
    \midrule
    Lack of reliability & 6 & 26\%\\
    Lack of scientific value & 3 & 13\%\\
    Lack of opportunity to use & 3 & 13\%\\
    Others & 3 & 13\%\\ 
    \bottomrule
  \end{tabular}
\end{minipage}
\end{table}

\textbf{\textit{Lack of reliability}} (6/23 occurrences, 26\%). This category was the main motivation that our respondents mentioned not to use GL in their research. This is related to the lack of rigor in which manner of GL content is written and published, putting into question the credibility of information presented due to the lack of quality control that makes it difficult to ensure their quality. Regarding this motivation, we present two quotes: \textit{``Because grey literature \ul{has no scientific or commercial control}, it can \ul{produce unreliable content with bias from the scientific point of view}''} and \textit{``GL is very open, \ul{without a deeper assessment of the material available}.''}


\textbf{\textit{Lack of scientific value}} (3/23 occurrences, 13\%). In this category, due to the lack of scientific value of GL by the scientific community, the respondents were afraid that GL use would weaken a research paper when submitted to the peer-review process, as a respondent cited: \textit{``Formally I never used it \ul{because I believe that will not be considered by academia. (...) academia only accepts peer-reviewed references}.''}

\textbf{\textit{Lack of opportunity to use}} (3/23 occurrences, 13\%). This category was mentioned due to the nature of research employed and GL is recent in the context of SE, as a respondent mentioned: \textit{``I never had an \ul{opportunity to use}.''} and the another mentioned: \textit{``I met this type of review recently and have not yet had the opportunity to adopt it in my research.''}



\textbf{\textit{Others}} (3/23 occurrences, 13\%). Here we group other responses that we were unable to group. Among them: 1) one that was removed from the entire analysis, where a researcher mentioned that s/he had never used GL because s/he never heard about GL before, showing that s/he didn't understand what the question asks for; and 2) another mentioned due to the lack of support for GL search. \textit{``Because I don't know \ul{where to search} for relevant content.''}


\begin{shaded}
\noindent
\textbf{Answer to RQ1:} Brazilian SE researchers are using GL motivated mainly to understand new topics, to find information about practical and technical questions, and to complement research findings. However, some researchers affirmed that avoid to use GL, in particular, as references in scientific papers.
\end{shaded}

\subsection*{RQ2: \rqtwo}

Here we intend to investigate the GL source used. To answer this question, we used the responses of the 53 respondents that mentioned use GL. When analyzing these questions, we found several sources that are used by SE researchers, as listed in Table~\ref{tab:sources-GL}. We highlight that 11 out of 53 (20.7\%) respondents mentioned the use of a search engine (e.g., Google, Google Scholar) as a start point to find GL content. However, we did not consider Google as a source of GL, although the respondents of our survey had considered. In the following, we present some of our findings.

\textbf{\textit{Community websites}} (16/53 occurrences, 30.2\%). The most common source used was the community website, i.e., websites in which the users can interact with others, e.g., creating content, posting comments, assess the content. Some researchers mentioned the use of Stack Overflow and Quora as a GL source, as mentioned by a respondent: \textit{``\ul{Communities} that bring together a variety of developers profile, such as \ul{Stack Overflow}.''}

\textbf{\textit{Blogs}} (15/53 occurrences, 28.3\%). The use of blogs as a source of GL was the second most common category found. A respondent used blogs from renowned practitioners, as s/he pointed out: \textit{``Sites or \ul{blogs} by well-known authors in a particular area.''} Another respondent mentioned the content of blogs derived from companies that produce a diversity of material and content of SE and software development in general: \textit{``\ul{Blogs} by SE firms (Netflix, Uber, Facebook engineering) (...).''}

\textbf{\textit{Technical experience/report/survey}} (14/53 occurrences, 26.4\%). Most of the respondents that mentioned this category used technical experience, reports, and surveys derived from industry, as a respondent pointed out: \textit{``Usually websites of companies that provide \ul{technical reports}, for instance, such as SEI, CMU, Jetbrains, among others.''} Another SE researcher mentioned that there are also technical reports derived from academic settings: \textit{``\ul{Technical Reports} published in national and international research groups, available on publications repositories.''}



\textbf{\textit{Companies website}} (8/53 occurrences, 15\%). This category means the website of companies, e.g., Google, Facebook, and ThoughtWorks, that contains information regarding their technologies, methods, practices, etc. Some respondents mentioned browsing these websites to find news about a specific technology to help decision making. Regarding this category, a SE researcher pointed out: ``\textit{I have always used blogs, and \ul{companies' website} to help me with decision making to select a specific software or tool to use.''}

\textbf{\textit{Others}} (3/53 occurrences, 5.7\%). This last category group responses that we were not able to group elsewhere, which include government publications, open data portal, and class material.

\begin{table}[tp]
\caption{Sources of GL used by SE researchers.} 
\label{tab:sources-GL}
\begin{minipage}{.4\textwidth}
  \centering
  \begin{tabular}{lcc}
    \toprule
    Source & \# & \%\\
    \midrule
    Community website & 16 & 30.2\%\\
    Blogs & 15 & 28.3\%\\
    Technical experience/report/survey & 14 & 26.4\%\\
    Companies website & 8 & 15\%\\
    Preprints & 5 & 9.4\%\\
    Books & 5 & 9.4\%\\
    Data repository & 4 & 7.5\%\\
    Videos & 3 & 5.7\%\\
    Non-scientific magazines & 3 & 5.7\%\\
    News & 2 & 3.8\%\\
    Others & 3 & 5.7\%\\ 
    \bottomrule
  \end{tabular}
\end{minipage}
\end{table}

\begin{shaded}
\noindent
\textbf{Answer to RQ2:} We found several GL sources used by Brazilian SE researchers. The most common sources are the content of community websites, blogs, technical experience/reports/surveys, and companies' websites.
\end{shaded}

\subsection*{RQ3: \rqthree}

With this research question, we explore the criteria of how the SE researchers assess the credibility of GL. They were asked in one open-ended question. In this research, we found 16 cases of mention in a general way to the criteria of GL source need to be trustable. Table~\ref{tab:credibility-GL} summarizes the results, and some of them are described in the following.


\textbf{\textit{Renowned authors}} (15/53 occurrences, 28.3\%). Some respondents mentioned the content of GL provided by renowned authors is an important criterion to assess its credibility. For instance, they assess the author's experience and reputation about the topic on the community, as some respondents cited Martin Fowler as an important software engineer with notorious knowledge. 
A respondent mentioned the importance of relying on a renowned author: \textit{``One source that \ul{shows an author with an in-depth knowledge} about they are writing.''} Another respondent mentioned the importance of searching practitioners: \textit{``popular blogs and websites \ul{from important people of the industry}.''}

\textbf{\textit{Renowned institutions}} (14/53 occurrences, 26.4\%). Similar to the above category, in this, we perceived that an important criterion of credibility is the GL use available by renowned institutions or research groups, as a respondent mentioned: \textit{``Something (GL) that is produced by an institution with credibility on the topic.''} Regarding this criterion, another researcher pointed out: \textit{``When one \ul{recognized institution is supporting} (whether reviewing, following up, etc.) the work. For instance, the technical reports produced by SEI or by the Institute of Fraunhofer, because their institutions following a scientific rigor and concerned with the production of the material.''} Still, the groups of research of an institution were mentioned: \textit{``\ul{Repositories of research group publications} with a history and reputation of conduct research on the topic.''}

\textbf{\textit{Cited by others}} (8/53 occurrences, 15\%). This category was mentioned to express those respondents that considered as a trusted source which one that was cited by others (studies or people). In this sense, for instance, a respondent affirmed:
\textit{``\ul{The ResearchGate shows the citations and recommendations of works by other researchers}, even some of them were not peer-reviewed.''} Still, another researcher affirmed: \textit{``A \ul{source of information attested by the community} that used certain information.''} 
This last mention refers to the Stack Overflow, in which the users can comment, ``up vote'', and ``down vote'' the answers.

\textbf{\textit{Renowned companies}} (7/53 occurrences, 13.2\%). Some respondents considered as a trusted GL source renowned software industries or portals, as mentioned by a respondent: \textit{``Sites or blogs of \ul{large software engineering companies} (Netflix, Uber, Facebook).''}


\begin{table}[H]
  \caption{Criteria to assess GL credibility.} 
  \label{tab:credibility-GL}
\begin{minipage}{.4\textwidth}
  \centering
  \begin{tabular}{lcc}
    \toprule
    Criteria & \# & \%\\
    \midrule
    Renowned authors & 15 & 28.3\%\\
    Renowned institutions & 14 & 26.4\%\\
    Cited by others & 8 & 15\%\\
    Renowned companies & 7 & 13.2\%\\
    \bottomrule
  \end{tabular}
\end{minipage}
\end{table}

\begin{shaded}
\noindent
\textbf{Answer to RQ3:} The major of the criteria of credibility is about who produces the content of GL, whether produced by a person, institution, company, etc., since the source is renowned.
\end{shaded}

\subsection*{RQ4: \rqfour}

Our last research question intends to explore the perceived benefits and challenges (problems or difficulties) on the GL use by SE researchers. They were asked in two open-ended questions. The results regarding the benefits are presented in Table~\ref{tab:benefits-GL} and the challenges in Table~\ref{tab:challenges-GL}. In the following, we present some discussions about our findings.

\begin{table}[b!]
\caption{Benefits on the use of GL.}
\label{tab:benefits-GL}
\begin{minipage}{.4\textwidth}
  \centering
  \begin{tabular}{lcc}
    \toprule
    Benefit & \# & \%\\
    \midrule
    Easy to access and read & 16 & 30.2\%\\
    Provide a Practical Evidence & 13 & 24.5\%\\
    Knowledge acquisition & 13 & 24.5\%\\
    Updated information & 6 & 11.3\%\\
    Advance the state of the art/practice & 5 & 9.4\%\\
    Different results from scientific studies & 3 & 5.7\%\\
    \bottomrule
  \end{tabular}
\end{minipage}
\end{table}

\subsection*{Benefits}

\textbf{\textit{Easy to access and read}} (16/53 occurrences, 30.2\%). This category was the most common benefit perceived by the respondents, mainly because most of GL sources are open access, are easily recovered by free search engines, and the contents are usually easy to read. A respondent mentioned the information in GL is written in a less formal language: \textit{``\ul{Easy to access} and \ul{is written in less formal language}.''} Other respondent shares the same opinion: \textit{``The content usually \ul{have easier access} and a \ul{more accessible language}.''}

\textbf{\textit{Practical evidence}} (13/53 occurrences, 24.5\%). Respondents mentioned that GL provides evidence from the industry, which is important to understand the state of the practice. A respondent mentioned that used GL to find information not found in traditional literature: \textit{``To discover \ul{practical information and practices} not reported on traditional literature.''} Another researcher shared the same opinion: \textit{``Understanding \ul{how things happen in the industry} (...).''}

\textbf{\textit{Knowledge acquisition}} (13/53 occurrences, 24.5\%). Some respondents mentioned if using only the traditional literature, the knowledge is limited, for this reason, the GL use could permit to widen the knowledge with different information, as a researcher mentioned: \textit{``The industry experience reports brought \ul{different facets about the phenomenon} they were studying.''} Another situation was pointed out by a respondent that read a researcher's blog: \textit{``(...) \ul{more complete and detailed data about one scientific research than scientific articles} of the same author.''}

\textbf{\textit{Updated information}} (6/53 occurrences, 11.3\%). Since it often takes a reasonable time to have a scientific paper published, the content of some papers may become technically outdated shortly after publication. In this sense, our respondents mentioned that GL is often more up-to-date when it comes to technical details. Regarding this situation, a respondent affirmed: \textit{``(...) Additionally, \ul{newer technologies tend to appear faster in GL}.''} Another one claimed: \textit{``I have found very interesting (blog) articles \ul{about new topics}.''}

\textbf{\textit{Advance the state of art/practice}} (5/53 occurrences, 9.4\%). Some respondents perceived the importance of GL to better understand the industry and to conduct research aiming to find important gaps in the practice. A respondent affirmed: \textit{``Understanding \ul{how things happen in the industry}, and which technologies derived from academia are in use. The GL also \ul{reveals many gaps and opportunities to applied research and to transfer of knowledge}.''}

\textbf{\textit{Different results from scientific studies}} (3/53 occurrences, 5.7\%). Some researchers revealed the importance of GL in providing additional knowledge not yet available in the research area. 
Regarding this benefit, a respondent pointed out: \textit{``\ul{Data and evidence (of GL) are different from peer-reviewed articles} that do not always provide original data for replication and also by limiting the coverage and comprehensiveness of data available from non-GL sources.''}

  \begin{table}[b!]
    \caption{Challenges on the use of GL.}
    \label{tab:challenges-GL}
    \begin{minipage}{.4\textwidth}
      \centering
      \begin{tabular}{lcc}
        \toprule
        Challenge & \# & \%\\
        \midrule
        Lack of reliability & 34 & 64.2\%\\
        Lack of scientific value & 15 & 28.3\%\\
        Difficult to search/find information & 6 & 11.3\%\\
        Non-structured information & 6 & 11.3\%\\
        \bottomrule
      \end{tabular}
    \end{minipage}
  \end{table}

\subsection*{Challenges}

\textbf{\textit{Lack of reliability}} (34/53 occurrences, 64.2\%). This category was the main challenge perceived by the respondents, some of them put in check the reliability of GL's content, as a researcher pointed out: \textit{``The biggest challenge, in my opinion, \ul{represents the validation of what is being reported}.''} Still, another pointed out: \textit{``How to \ul{ensure the quality of information} maybe is the big challenge to use GL.''}

\textbf{\textit{Lack of scientific value}} (15/53 occurrences, 28.3\%). This was the second category most cited by the respondents. This category is closely related to the one mentioned before. Some respondents cited this problem because they are not comfortable to use GL due to the lack of recognition of this source by scientific area or to use this source as a reference in scientific work, as two respondents affirmed: \textit{``It has \ul{not scientific rigor}''}, and \textit{``(...) The diversity of channels that they are published hinder the search, \ul{defy replicability} (...).''}

\textbf{\textit{Difficult to search/find information}} (6/53 occurrences, 11.3\%). The diversity of sources to search for GL content was a challenge perceived for some respondents, as pointed out: \textit{``The \ul{diversity of channels} in which the content of GL are published hinder the search, defy replicability, and \ul{increase the effort to select content}.''} 


\textbf{\textit{Non-structured information}} (6/53 occurrences, 11.3\%). Another challenge perceived is the content structure of a GL source. For instance, to some respondents, there is a lack of a writing pattern and a large variety of formats in which the sources are published. Regarding those challenges, a respondent mentioned: \textit{``The \ul{lack of pattern} to the structure and writing''}, and another complement: \textit{``The \ul{variety of formats in which the sources (non-standard)} are reported in GL also configured as another significant challenge.''}

\begin{shaded}
\noindent
\textbf{Answer to RQ4:} We found several benefits, the most common was the content of GL is easy to access and read, and this is important to knowledge acquisition, mainly about providing practical evidence derived from SE practitioners. Regarding the challenges, the most cited were related to the difficulty to use GL in scientific research, due to the lack of reliability and lack of scientific value. 
\end{shaded}

\section{Discussion}\label{sec:discussion}



\subsection{Revisiting findings}
Even though several benefits and challenges were perceived, some of them seem contradictory. In fact, they are part of the trade-off between white literature and GL natures. For instance, on the one hand, it is \textit{Easy to access and read} the content of GL. On the other hand, it is \textit{Difficult to search/find information} due to sources' variety. We noticed that when the respondent mentioned the benefit, they answered about the access of the GL source that is not restricted as most of the scientific papers. Regarding the content, because a GL content is usually written in an informal language. However, this challenge may arise when they think about how to retrieve information, for instance, automatically, that is not easy due to the diversity of content and the \textit{Non-structured information}.

Another important trade-off is the benefit \textit{Advance the state of the art/practice} and the challenges \textit{Lack of reliability} and \textit{Lack of scientific value}. Those exacerbate some of the need for attention even with the perceived benefit, several researchers avoid the use of GL due to those challenges. Those trade-offs were expected, in part, but they also show the need for further investigation on how to improve the content provided and to better deal with them. For this reason, we claimed in our lessons learned the necessity to improve them.

Important findings of criteria to assess the GL credibility showed that most of them are related to the producer of content be renowned (authors, institutions, and companies). It caught our attention that no mention was done on how to assess the content of a GL, despite the challenge \textit{Lack of reliability} that is related to this. This leads us to question whether to assess the credibility of a GL source being a recognized source is sufficient, without even evaluating the trust of content.

Even we confirmed some findings of the literature, our main category of benefit \textit{Easy to access and read} was not mentioned by previous studies. It is important to emphasize that our study counts the number of times where a category was found, aimed to show the strength of each one.


\subsection{Lessons learned}
With this study and knowledge about previous related work, we claim to the potential of the GL to SE research and practice. However, some important advice is needed, both to SE researchers and practitioners.

\vspace{0.2cm}
\noindent
\textbf{Researchers:} Our findings highlight to pay attention when searching, selecting, and using grey literature in their research: 1) explore the GL sources before using on their research because there are several types of GL source. This should aim to understand how to retrieve information from them, due to the issues about the difficulty to search for; 2) select data produced by a renowned source (e.g., SEI, Facebook) aiming to increase the credibility of the scientific potential. Still, it is essential to add some criteria to assess the data content; and 3) understand how to improve the search for GL using a systematic approach with methods and techniques to better deal with the content, aiming to reduce their lack of reliability.

\vspace{0.2cm}
\noindent
\textbf{Practitioners:} Our findings show the importance of the content provided by them for the research community. However, for this information to be consumed by researchers and to create a relevant impact on academia, we include some advice for practitioners: 1) substantiate the data presented in an accessible language and with detail information, e.g., explaining the context, making the used data available; 2) adopting some quality criteria to improve the credibility of their content, e.g., use a checklist to verify if the information is well described; and 3) adopting a pattern to provide information makes easier to retrieve information using an automatic approach. We understand the third piece of advice is a gap at the moment, however, it raises future work possibilities. Cartaxo et al.~\cite{Cartaxo:2016:ESEM} propose the use of evidence briefings to describe findings for practitioners and is an example that can be used.



\subsection{Limitations}
\textbf{Construct validity:} During our process to draw our questionnaire, before sending the actual survey, a draft survey was reviewed by an experienced researcher. After, we evaluated our survey design conducting pilot studies with two SE researchers. Even our efforts to elaborate our questionnaire, some bias may have occurred, for instance, the definition used of GL was broad, which made it impossible for our findings to be more specific to understand the GL sources used and their criteria to assess their credibility.

\vspace{0.2cm}
\noindent
\textbf{Internal validity:} As occurred in any qualitative research, some subjective decisions with personal interpretation may have occurred during the data extraction and analysis of the survey responses. Aiming to minimize those biases, we used a peer-review approach, and we invoke a third researcher to revise the derived codes.

\vspace{0.2cm}
\noindent
\textbf{External validity:} In our research, we conducted our survey in the largest SE conference in Brazil and was collected answers from Brazilian SE researchers. We believe our sample is representative of SE research because we had a 30\% response rate with a diversity of respondents (1/3 are women, ~50\% have a Ph.D. in SE and ~30\% a Master's). However, we can not guarantee that the rest of the respondents have previous experience with research. Moreover, as we focused on the Brazilian SE research community; the findings may not apply to other populations. Although, we used the peer review process during all this research aiming to improve the external validity to draw general conclusions.

\vspace{0.2cm}
\noindent
\textbf{Conclusion validity:} Even our ~30\% response, it is possible that some important information was missed. However, we compared our results with previous studies conducted with different populations and our results showed similarly.

\section{Related Work}\label{sec:related_works}



\subsection*{The use of GL in primary studies} SE researchers are relying on several sources of GL to answer their research questions. Some examples include screencasts, YouTube, Twitter, and Stack Overflow. In the following, we briefly describe some of these studies. MacLeod et al.~\cite{MacLeod:2015:ICPC} conducted  studies exploring the use of screencasts, for instance, they investigated how and why developers create and share screencasts through YouTube. Some motivations (e.g., learning, 
self-promotion) and a diversity of goals and techniques for creating such screencasts (e.g., code demonstrations, describing code functionality in different ways) were found. 
Some researchers investigated the use of Twitter in SE as an important social media for keeping up with new technologies and the fast-paced development landscape~\cite{Storey:2017:TSE,Singer:2014:Twitter}. Twitter was also associated with communicating issues, documentation, to advertise blog posts to its community, as well as to solicit contributions from users~\cite{Storey:2014:ESM}. 
Other researchers have investigated the Questions and Answers (Q\&A) websites, for instance, Zahedi et al.~\cite{Zahedi:2020:EASE} employed an empirical study aimed at exploring Continuous Software Engineering (CSE) from the practitioners' perspective by analyzing 12,989 questions and answers from Stack Overflow. The findings present trends (questions are becoming more specific to technologies and more difficult to attract answers), and the most challenging areas in this domain form the practitioners' perspective.

\subsection*{How researchers use GL?}
Garousi et al.~\cite{Garousi:2016:EASE} investigated the potential use of GL in SLR comparing the results in which the was included the GL as primary study and the other not. The findings showed that with GL, the results could be useful to answer practical and technical research questions, bringing the results closer to SE practice.
Raulamo-Jurvanen et al.~\cite{Raulamo-Jurvanen:2017:EASE} conducted the first Grey Literature Review (GLR) we have known in SE to analyze how software practitioners address the practical problem of choosing the right test automation tool. The data was derived from the experiences and opinions present in most of the findings. Moreover, this research examined the evidence available at the GL sources to add the credibility of the claims of their content, for example, the number of readers and sharing, number of comments, number of Google Hits for the title and the analyzes for the sources were backlinks (a reference comparable to a citation). Another GLR we found was about pains and gains of the use of microservices~\cite{Soldani:2018:GLR}. In this study, it was observed that, in traditional literature, academic research on the topic is still in its early stage even though companies are working day-by-day with microservices, as also witnessed by the considerable amount of GL on the topic.

Neto et al.~\cite{Neto:2019:ESEM} conducted the first tertiary study that focused only on Multivocal Literature Review (MLR) and Grey Literature Review (GLR), with the aim to provide preliminary work about the current research involving GL. Were selected 12 studies (ten using MLR and two using GLR) in which were explored their motivations to included GL. The preliminary findings showed that the lack of academic research on the topic, emerging research on the topic, and to complement evidence with the GL were the main reasons.

Williams and Rainer conducted three studies aimed to understand the use of blog articles in SE research. The first study~\cite{Williams:2017:EASE} examined some criteria to evaluate blog articles to be used as a source of SE research evidence through two pilot studies (a systematic mapping study and preliminary analyses of blog posts). The findings showed some criteria to select the content (e.g., authentic, informative) 
of a blog article. Some benefits (e.g., evidence timeliness, trends analysis) and drawbacks (content diversity) using blogs as an evidence source in SE research were also found.
The second study~\cite{Williams:2018:ASWEC} informally reviewed how practitioners use blogs, review the research literature, and present the findings of a survey. An overview of research on this topic was presented, exploring some potential benefits (e.g., trend analysis, practitioners insights evidence) and challenges (e.g., the variability of blog content, un-established process for assessing the quality).
The third study~\cite{Williams:2019:EASE} focused on finding credibility criteria to assess blog posts by selecting 88 candidate criteria of credibility from a previous Mapping Study~\cite{Williams:2017:EASE}. Then, were surveyed 43 SE researchers to gather opinions on a blog post to assess those credibility criteria. Some criteria were found, for instance, the presence of reasoning, reporting empirical data, and reporting data collection methods.

Most recently, Zhang et al.~\cite{Zhang:2020:ICSE}
investigated GL in two perspectives: 1) conducted a tertiary study to identify Secondary studies that used the term ``grey'' or ``multivocal'' in their studies, aiming to understand the definitions of GL used by researchers, and the types of GL used; 2) surveyed with 35 SE researchers of included secondary studies and invited SE experts to understand the motivations and challenges to use GL, how they used GL in their studies, and how they search for it. 

Even though the similarity of these works with our work, there are differences in at least five points: 1) we did not focus on a specific type of GL source; 2) we explored the experience of SE researchers to understand which type of GL they have used; 3) we tried to understand what motivates and demotivates SE researchers to use GL; 4) we found different criteria to assess GL credibility; and 5) we explored a broader population of SE researchers, not only experimental SE researchers.

Our study confirmed some findings of previous studies (e.g., the benefits of GL provides updated information~\cite{Williams:2017:EASE} and different results of scientific studies~\cite{Williams:2018:ASWEC}, and the challenges of lack of reliability~\cite{Zhang:2020:ICSE} and non-structured information~\cite{Williams:2018:ASWEC}), showing the importance of GL for the SE research area. However, some of our findings differ from them because we investigated some area that has not been explored, such as, we do not focus on a specific GL source and we aimed to understand the motivations to use and reasons to avoid a GL in a specific SE research community. Still, we found some findings not mentioned in previous studies
~\cite{Williams:2017:EASE,Rainer:2018:TR,Williams:2019:EASE,Zhang:2020:ICSE}: 1) our most common benefit \textit{Easy to access and read} and the second most common category of challenges \textit{Lack of scientific value}; and 2) two credibility criteria, the \textit{Renowned institutions} and \textit{Renowned companies}.



\section{Conclusions and Future Work}\label{sec:conclusions}
Grey Literature stands as an important source to SE research and practice since SE practitioners rely upon and use social media communication channels to interact and share their thoughts and data about their experiences and projects. For this reason, in the last years, several studies have explored the content provided by GL source and others investigating how researchers use them.

In this work, we conducted the first investigation of GL use from the Brazilian SE researchers' perspective we have known, to better understand GL sources usage, potential benefits and challenges, and criteria to assess GL credibility. Our major findings show: 1) there are several motivations to Brazilian SE researchers use GL, mainly because its content provides important information to researchers. However, it is still hard to find reliable information for scientific research; 2) several GL sources are being used by Brazilian SE researchers. The most common was blogs, community websites, and technical experience/reports; 3) some criteria to assess GL credibility are renowned people, institutions, and companies responsible for the content; and 4) a diversity of benefits and challenges using GL were perceived by SE researchers. Regarding benefits, we found the content of GL is easy to access and read, it provides practical evidence, and it is important to knowledge acquisition. Some challenges we also found are mainly about the lack of reliability and scientific value. 
The findings of this research showed that even the potential of GL, some trade-off may arise that need the attention of investigation to make the use of GL more mature, something common to happen as it is a new and growing area in SE.

For \textit{future works}, we plan to: 1) conduct a large scale study about GL in SE to expand our sample to other SE research communities; 2) investigate a set of criteria to improve the assessment of the credibility of GL; 3) to provide a guideline on how to search and find information of GL; 4) investigate on how to assess and retrieve valuable information to increase the scientific value of GL; and 5) to investigate and provide a guideline to SE practitioners to make their content valuable to research.

\section{Acknowledgments}
We thank the anonymous reviewers and the SE researchers for participating in the study. This work is partially supported by INES 2.0 (www.ines.org.br), CNPq grant 465614/2014-0, FACEPE grants APQ-0399-1.03/17 and APQ/0388-1.03/14, CAPES grant 88887.136410/2017-00. Sérgio Soares is partially supported by CNPq grant 309697/2019-0.

\bibliographystyle{ACM-Reference-Format}
\bibliography{references}
\end{document}